\documentstyle[aps,epsf]{revtex}
\def\be{\begin{equation}}
\def\ee{\end{equation}}
\def\bea{\begin{eqnarray}}
\def\eea{\end{eqnarray}}
\def\ds{\displaystyle}

\newtheorem{Result}{Result}
\begin{document}
\title{The role of anisotropy and inhomogeneity in Lemaitre--Tolman--Bondi collapse}
\author{Filipe C. Mena$^{1,3}$, Brien C. Nolan$^2$ \& Reza Tavakol$^3$\\
{\em $^1$
Departamento de Matem\'atica,
Universidade do Minho, Campus de Gualtar,
4710 Braga, Portugal}
\\
{\em $^2$
School of Mathematical Sciences,
Dublin City University, Glasnevin, Dublin 9,
Ireland}
\\
{\em $^3$
Astronomy Unit, School of Mathematical Sciences,
Queen Mary, University of London,
London E1 4NS, U.K.}}
\date{November, 2003}
\maketitle
\begin{abstract}
We study the effects of shear and density inhomogeneities
in the formation of naked singularities in spherically symmetric dust space--times.
We find that in general neither of these physical features alone uniquely specifies the
end state of the gravitational collapse.
We do this by (i) showing that, for open sets of initial data,
the same initial shear (or initial density contrast) 
can give rise to both naked and covered solutions.
In particular
this can happen for zero initial shear
or zero initial density contrast;
(ii) demonstrating that both shear and density contrast are invariant under
a one parameter set of linear transformations
acting on the initial data set and 
(iii) showing that asymptotically (near the singularities)
one cannot in general establish a direct relationship between the
rate of change of shear (or density contrast)
and the nature of the singularities.
However, one can uniquely determine the nature of the singularity if both
the initial shear and initial density contrast are known.

These results are important in understanding the effects of the
initial physical state and in particular the role of shear,
in determiming the end state
of the gravitational collapse.
\end{abstract}
\section{Introduction}
An important outcome of general relativity has been to show that
subject to a number of physically reasonable assumptions, the
final state of gravitational collapse will be singular in a range
of settings (see e.g. \cite{Hawking-Ellis73}). An
outstanding question concerns
the nature of the resulting singularities and in particular
whether and under what conditions they may be naked or black
holes \cite{Penrose,Wald97}. 
Over the recent years a great deal of effort has gone into the
study of these questions. These have mainly concentrated on
collapse in spherically symmetric settings and involve the study
of collapse of scalar field \cite{chris99,Gundlach} as well as
other matter sources
including dust \cite{Dwivedi-Joshi97,MTJ}, perfect fluids
\cite{Giambo-etal03-1,Harada02}, imperfect fluids 
\cite{Harada99,Giambo-etal03-2,Sergio02,Joshi-Goswani} and null strange quark fluids
\cite{Dadhich,Harko}.   

All these studies show that the end state of the spherical
collapse can be either a black hole or a naked singularity, depending upon the nature
of the initial data and the kinematical properties near the singularity. 
Apart from the mathematical questions concerning the likelihood of
each end--state, an important question from a physical point of
view is how the nature of the end--state of collapse in the
spherical settings may depend on physical characteristics of the
regime under consideration. In particular, we ask how 
does the inhomogeneity and anisotropy influence the process and
the final outcome
of collapse?

This issue has been studied by a number of
authors.
Penrose \cite{Penrose79} discusses the possibility that anisotropy might play a role
in the occurence of naked singularities.
Herrera et al. \cite{Herrera}
considered a spherically symmetric collapsing space--time with an anisotropic
fluid
(with radial and tangential
pressures) and
showed
that high density contrasts increase the radial
velocity of collapse. They also proved that the local anisotropy of the
pressure might have the same effect
suggesting that this might lead to naked singularities (depending on some
inequalities between radial and tangential pressures).

More recently, Joshi, Dadhich and Maartens \cite{JDM} have also
considered this question in the special case of marginally bounded
Lemaitre-Tolman-Bondi
(LTB) collapse and found that the shear associated with the initial
distribution of matter fully determines the final outcome of collapse in
these settings. In particular, they have found that 
the stronger the shear near the singularity the more
likely it is to have a naked singularity.
Soon after, Chan et al. \cite{Chan} found a self-similar solution with
heat--flow
which
was shear-free and, 
nevertheless, exhibited naked singularities. 

Here, motivated by these previous results, we make a
detailed study of the relation between the nature of the
end--state of LTB collapse with anisotropy and
inhomogeneity, by considering more general settings.
We study how the choices of 
shear and density contrast functions can influence the 
nature of the gravitational singularities.
In particular, we consider the behaviour of the 
shear function initially, asymptotically (near the
singularity) as well as at intermediate times 
and study the effects of each on determining the end state of
gravitational collapse.

The structure of the paper is as follows: In section 2 we give a
brief description of the spherical dust models. In Sections 3 and 4
we consider the effects of the shear on
the end state of collapse. In section 5 we consider in turn the density
contrast. Section 6 contains some remarks about space-time extensions
after shell-crossing and initial data sets. Finally, we conclude in Section 7.
\section{Spherically symmetric dust space--times}
The inhomogeneous spherically symmetric dust space--time can
be represented by the Lemaitre-Tolman-Bondi (LTB)
line element \cite{Lemaitre,Tolman,Bondi}
\be
\label{metric-tolman}
ds^2 =
 -dt^2 + \frac{{R^{'}}^2}{1+E} dr^2 +R^2 (d\theta^2 + \sin^2
\theta
d\phi^2),
\ee
where $r, \theta, \phi$ are comoving coordinates.
The dot and prime denote differentiation with
respect to $t$ and $r$ respectively and
$R=R (r,t)$ and $E=E(r)$ are $C^2$ real functions
such that $R (r,t)\ge 0$ and $E(r)>-1$.
The matter--density is given by
\be
\label{density}
\rho(t,r) = \frac{m'}{R^2 R^{'}}
\ee
where $m=m(r)$ is another $C^2$ real function such that $m(r)>0$.
The evolution equation for the case of $\dot{R}<0$
(corresponding to gravitational collapse)
takes the form
\be
\label{genfried}
\dot R = - \sqrt{\frac{m}{R} + E},
\ee
and can be solved
for different values of $E$
in the following
parametric forms:
\\

\noindent {For $E<0$}:
\bea
\label{elliptic}
&R =  {\ds \frac{m}{2(-E)}(1-\cos\eta)} \nonumber  \\
&(\eta-\sin\eta)  =  {\ds \frac{2(-E)^\frac{3}{2}}{m}}(t_c-t)
\eea
where $0< \eta < 2\pi$ and $t_c=t_c(r)$ is a third $C^2$ real function that
corresponds to
the time of arrival of each shell $r$
to the central singularity. Note that here and below, $t=t_c(r)$ corresponds to $\eta=0$. 
\\

\noindent {For $E>0$}:
\bea
\label{hyperbolic}
&R = {\ds \frac{m}{2E}(\cosh\eta-1)}, \nonumber  \\
&(\sinh\eta-\eta) = {\ds \frac{2E^\frac{3}{2}}{m}}(t_c-t)
\eea
where $\eta > 0$.
\\

\noindent {For $E=0$}:
\be
\label{parabolic}
R=\left(\frac{9m}{4}\right)^\frac{1}{3}(t_c-t)^\frac{2}{3}.
\ee
In what follows we find it also useful to work with the compact form for the
solution
to equation (\ref{genfried});
\be
\label{evolution}
t_c-t=\sqrt{\frac{R^3}{m}}G(-{\ds\frac{ER}{m}}),
\ee
where
$G$ is a positive real function given by
\bea
\label{Gfunctions}
&G(x)=&\frac{arcsin(\sqrt{x})}{x^{3/2}}-\frac{\sqrt{1-x}}{x},~~~~~~~~
for~~~1\ge x>0\nonumber\\
&G(x)=&\frac{2}{3},~~~~~~~~~~~~~~~~~~~~~~~~~~~~~~~~~~~ for~~~ x=0\\
&G(x)=&\frac{-arcsinh(\sqrt{-x})}{(-x)^{3/2}}-\frac{\sqrt{1-x}}{x},
~for~~~ 0> x > -\infty.
\nonumber
\eea
Using the coordinate freedom to rescale
\be
\label{rescale}
R(0,r)=r,
\ee
at an initial time $t=0$,
equation (\ref{evolution}) gives
\be
\label{tnot}
t_c(r)=\frac{r^{3/2}G(p)}{\sqrt{m}},
\ee
where $p=-rE/m$. So, the initial data set is given by ${\cal I}=\{m(r),E(r)\}$.

We note that the metric (\ref{metric-tolman}) can be matched at
a boundary, say $r=r_d=const.$, to the Schwarzschild metric in
the exterior region. Thus the scenario here is
that of a collapsing compact matter region matched to an exterior
Schwarzschild space--time.

We shall refer to a singularity as naked if there is a family of
future directed non--spacelike geodesics which terminate at the
singularity in the past. In spherical symmetry, if there are no
radial--null geodesics emerging from a central singularity then the singularity is necessarily
censored \cite{NMG}. Therefore
we shall only be concerned with outgoing radial null geodesics which, as can
be seen from (\ref{metric-tolman}), correspond to the solution of
the differential equation 
\be 
\label{geo}
\frac{dt}{dr}=\frac{R'}{\sqrt{1+E}}. 
\ee
One can rewrite this
equation as (see \cite{Dwivedi-Joshi97}) 
\be
\label{geod}
\frac{dR}{du}=\frac{1}{u'}\left(R'+\dot{R}\frac{dt}{dr}\right)=\left(1-
\sqrt{\frac{E+\Lambda/X}{1+E}}\right)H(X,u), 
\ee
where
\bea
&H(X,u)&=
(\eta_u-\beta_u)X+\\
&&\left(\Theta_u-
(\eta_u-\frac{3}{2}\beta_u)X^{\frac{3}{2}}G(-PX)\right)\sqrt{P+\frac{1}{X}}\nonumber
\eea
and
\bea
&&X=\frac{R}{u},~\eta= r\frac{m'}{m},~\eta_u= \eta \frac{u}{ru'},~
\beta=r\frac{E'}{E},~\beta_u=\beta \frac{u}{ru'},\nonumber\\
&&P=\frac{uE}{m},
~p=-r\frac{E}{m},~\Theta_u=\Theta\frac{\sqrt{r}}{\sqrt{u}u'},~
\Lambda=\frac{m}{u},\\
&&\Theta=\frac{\sqrt{m}}{\sqrt{r}}t'_c(r)=
\frac{1+\beta-\eta}{\sqrt{1-p}}+(\eta-\frac{3}{2}\beta)G(p),\nonumber
\eea
with the positive real function $u=u(r)$ being monotonically increasing and such
that $u(0)=0$. Later on
we will specify $u(r)=r^3$.
In the cases where $E(r)=0$ we will take $\beta(r)=0$.
Following \cite{Dwivedi-Joshi97}, we use the
algebraic equation in $X_0$ (obtained from (\ref{geod})) 
\be 
\label{general}
\left(1-\sqrt{\frac{E_0+\Lambda_0
/X_0}{1+E_0}}\right)H(X_0,0)-X_0=0, 
\ee 
where the subscript $`0`$ denotes the limit of the associated
functions as $r\to 0$ (respectively $u\to 0$) along the geodesic,
in order to demonstrate the existence of radial-null 
geodesics emanating from the singularity
in spherical symmetric dust collapse.

An important outcome of past studies (e.g. \cite{Dwivedi-Joshi97})
is that the occurrence of black holes or naked singularities as
final outcomes of collapse depend on the choice of initial data
$\{m(r),E(r)\}$.
This initial data set has to satisfy certain {\em regularity conditions}
namely: 
\begin{enumerate}
\label{cond}
\item there are no trapped surfaces initially, i.e. $R(0,r)>m(r)$, for
$r>0$.
\item there is no shell--crossing during
collapse, i.e. $R'(t,r)\ne 0$, for $r>0$. 
\item the initial matter--density $\rho(0,r)$ is non-zero at the center $r=0$.
\end{enumerate}
\section{Dependence of the nature of singularity on shear $\sigma(t,r)$}
In this section we study the shear $\sigma(t,r)$, in particular its behaviour 
in the approach to the singularity as well as 
the effects of changing $\sigma(t,r)$ on the nature of the singularity.

For the LTB metric the shear function (relative to the fluid congruence given
by the $4-$velocity vector field) is given by 
\be
\sigma(t,r)=\frac{\sqrt{3}}{3}\left(\frac{\dot{R}}{R}-\frac{\dot{R}'}{R'}\right)
\ee 
and its time derivative by 
\be 
\label{sigmadot}
\dot{\sigma}(t,r)=\frac{\sqrt{3}}{3}\left[\frac{\ddot{R}}{R}-\left(\frac{\dot{R}}{R}\right)^2-
\frac{\ddot{R}'}{R'}+\left(\frac{\dot{R}'}{R'}\right)^2\right].
\ee 
We start by recalling the following well--known result (see e.g. \cite{MTJ}):
\begin{Result}
Consider a dust spherically symmetric collapsing space--time satisfying the
regularity conditions. If
$\sigma(t,r)=0$ then the final state of collapse is a covered singularity.
\end{Result}
To proceed, we shall use the parametric solutions of the evolution equation
to calculate $\dot R/R$ and $\dot R'/R'$ in cases $E<0$ and $E>0$.
 The case $E=0$ has already been considered
in \cite{JDM}.
\subsection{CASE $E<0$}
In this case we have
\be 
\dot R=\sqrt{-E}\frac{\sin\eta}{\cos\eta-1},
\ee 
together with 
\be
R'=\left(\frac{m}{2(-E)}\right)'(1-\cos\eta)+\frac{m}{2(-E)}\sin\eta\frac{\partial\eta}{\partial
r} 
\ee 
and
\be
\dot
R'=(\sqrt{-E})'\frac{\sin\eta}{\cos\eta-1}-\sqrt{-E}\frac{1}{\cos\eta-1}\frac{\partial
\eta}{\partial r},
\ee
where ${\ds \frac{\partial \eta}{\partial r}}$ is given in the 
Appendix. We can then write 
\be
\sigma(t,r)=\frac{\sqrt{3}}{3}\left(\frac{2(-E)^{3/2}}{m} \right)
\left(\frac{-\sin\eta}{(\cos\eta-1)^2}+
\frac
{
\frac{E'}{2E}\sin\eta-\frac{\partial\eta}{\partial r}
}
{
\left(\frac{E'}{E}-\frac{m'}{m}\right)
(\cos\eta-1)^2+\sin\eta(\cos\eta-1)\frac{\partial \eta}{\partial r}
}
\right).
\ee
It can be seen that $\sigma(t,r)$ is invariant under the transformation
\begin{eqnarray} \Phi_a:{\cal{I}}&\to& {\cal{I}}\\
(m,E)& \mapsto& (a^{3/2}m,aE),
\end{eqnarray}
for an arbirtrary free parameter $a>0$. Note that $t_c(r)$ 
does change under this transformation, and so it is a non-trivial mapping on ${\cal I}$, i.e.
it leads to a different LTB space--time.
So, the same shear function
$\sigma(t,r)$ can correspond to different initial data functions $m$ and $E$.

Since we are mainly interested in the behaviour of $\sigma(t,r)$ around the singularity we use 
an expansion around $\eta=0$ to obtain 
\be
\sigma(t,r)=
-\frac{2\sqrt{3}(-E)^{3/2}}{3m}\left(\frac{6}{\eta^3}+\frac{1}{2\eta}+\frac{-2(AB)'+\frac{2}{3}AB'-BA'}{AB^2t_c'}
-\frac{7}{120}\eta\right)+O(\eta^2),
\ee
where
$A={\ds\frac{m}{2(-E)}}$ and $B={\ds\frac{2(-E)^{\frac{3}{2}}}{m}}$.
\subsection{CASE $E>0$}
In this case, we have 
\be
\dot R=\sqrt{E}\frac{\sinh \eta}{1-\cosh\eta},
\ee
together with
\be
R'=\left(\frac{m}{2E}\right)'(\cosh\eta-1)+\frac{m}{2E}\sinh\eta\frac{\partial\eta}{\partial
r},
\ee
and
\be
\dot
R'=(\sqrt{E})'\frac{\sinh\eta}{1-\cosh\eta}-\sqrt{E}\frac{1}{1-\cosh\eta}\frac{\partial\eta}{\partial
r},
\ee
where ${\ds \frac{\partial \eta}{\partial r}}$
is given in the 
Appendix.
We can then write 
\be
\sigma(\eta,r)=\frac{\sqrt{3}}{3}\left( \frac{2E^{3/2}}{m} \right)
\left(\frac{-\sinh\eta}{(\cosh\eta-1)^2}+
\frac
{
\frac{E'}{2E}\sinh\eta-\frac{\partial\eta}{\partial r}
}
{
\left(\frac{E'}{E}-\frac{m'}{m}\right)
(1-\cosh\eta)^2+\sinh\eta(1-\cosh\eta)\frac{\partial \eta}{\partial r}
}
\right).
\ee
We note that again $\sigma(t,r)$ is invariant under the transformation
$\Phi_a(m,E)=(a^{3/2}m,aE)$, for an arbitrary free parameter $a>0$. As a result, the same
shear function
$\sigma(t,r)$ can correspond to different initial data functions $m$ and $E$.
We use the above expressions and their expansions around $\eta=0$
to get
\be
\sigma(t,r)= -\frac{2\sqrt{3}E^{\frac{3}{2}}}{3m}
\left(\frac{6}{\eta^3}-\frac{1}{2\eta}+
\frac{-2(AB)'-BA'}{AB^2t_c'}+\frac{7\eta}{120}\right)+O(\eta^2),
\ee
where $A={\ds \frac{m}{2E}}$ and $B={\ds\frac{2E^{\frac{3}{2}}}{m}}$.
\subsection{Discussion}
\label{res}
The above results show that the shear $\sigma(t,r)$ for both $E>0$ and $E<0$ cases is
invariant under 
the transformation $\Phi_a(m,E)=(a^{3/2}m,aE)$, with $a>0$. This is not true for $\Theta_u$. 
The fact that $\Theta_{u}$ 
is not invariant under $\Phi_a$ has the important consequence of
allowing the
same shear function to produce both naked singularities and black holes. As a simple
example
we take $m(r)=m_3r^3+m_6r^6$ and $E(r)=E_2r^2+E_5r^5$.
In this case, letting $u=r^3$ 
we obtain from (\ref{general}) the quartic equation 
\be
\label{quadric-m3m6} 
2X_0^4+\sqrt{m_3}X_0^3-3\Theta_u
X_0+3\sqrt{m_3}\Theta_u=0, 
\ee 
with
\be
\label{theta}
\Theta_{u_0}= \frac{E_5/E_2-m_6/m_3}{\sqrt{1-p_0}}+
\left(\frac{m_6}{m_3}-\frac{3}{2}\frac{E_5}{E_2}\right)G(p_0).
\ee
Taking
$m_3=0.1,m_6=-0.9,E_2=-0.05,E_5=1$ we obtain positive roots for (\ref{quadric-m3m6})
which correspond to naked
singularities. 
Applying $\Phi_a$ with $a=0.61$ to the previous initial
data, 
we get black hole solutions instead and yet the same shear function $\sigma(t,r)$.   
Thus shear $\sigma(t,r)$ does not uniquely determine the final outcome of collapse.    

In the approach to the singularity, 
the shear in both $E>0$ and $E<0$ cases can be approximated by 
\be
\label{sigma-approx}
\sigma(t,r)=
-\frac{\sqrt{3}}{3}\frac{2|E|^{\frac{3}{2}}}{m}\left(\frac{6}{\eta^3}\right)+O(\eta^{-1}).
\ee
This demonstrates that the behaviour of $\sigma(t,r)$ near the singularity
for $E\ne 0$ 
depends  in general on both $m$ and $E$, whereas in the $E=0$ case
it depends only on $m$.
 
We can now study the shear strenght in the approach to the singularity
depending upon the values of 
$|E|^{\frac{3}{2}}/m$ as well as the relationship between $|E|^{\frac{3}{2}}/m$
and the time of horizon formation $t_H$.
The surface $t=t_H(r)$ is defined by $R=m$.
Using (\ref{evolution}), this gives 
\be 
t_H = t_c-mG(-E).
\ee
Using the expression (\ref{tnot}) for $t_c$, we have
\be
\frac{\partial t_H}{\partial m} = -\frac12
\frac{r^3}{m^{3/2}}(G(p)+2pG^\prime(p))-G(-E),
\ee
where $p=-rE/m$. Now $G$ is positive, and it turns out that $G(p)+2p
G'(p)$ is also positive, and so 
\be
\label{delay}
\frac{\partial t_H}{\partial m}<0.
\ee
Thus
for fixed $E$, regardless of sign, $t_H$ does increase with
decreasing $m$.


It is known that delays in the time of horizon formation can be associated
with the increased likelihood of finding naked solutions. From (\ref{delay}) 
we find that smaller $m$ produces delays in $t_H$.
Since smaller $m$ also induces stronger shear
around the singularity (see equation (\ref{sigma-approx})) one may be
tempted to relate stronger shears with delays
in $t_H$ and consequently with increased likelihood of finding naked
solutions, as was done in \cite{JDM} for $E=0$.
However, for $E\ne 0$, this is not necessarily true since in general one needs
both $m$ and $E$ to decide whether a solution is naked or not. 

Now recall that for LTB, given an initial function $m$ one can always
find a function $E$ such that a naked singularity solution exists
\cite{Dwivedi-Joshi97,Say}.   
So we may ask whether smaller $m$ values make it
easier to find $E$ functions                
which give rise to naked singularities.
In order to answer this question we also note that the part of the singularity
surface $t_c(r)$ which can produce naked singularities is $t_c(0)$, so we are mainly 
interested in the behaviour of $\sigma(t,r)$ around $r=0$.
 
We take $C^{\infty}$ functions $m$ and $E$ 
and integers $N>2, M>3$ so that we can always write
$m(r)=\sum^{N}_{i=3}
m_i r^i$, with $m_3\ne 0$ and $E(r)=\sum^{M}_{j=2}E_jr^j$, with $E_2\ne 0$. 
In this case, we get around $r=0$
\be
\label{cinfty}
\sigma(t,r)=-\frac{\sqrt{3}}{3}\frac{2|E_2|^{\frac{3}{2}}}{m_3}\left(\frac{6}{\eta^3}\right)+O(\eta^{-1}),
\ee
and we find naked singularity
solutions from
equation (\ref{general}) if and only if (see e.g.
\cite{Dwivedi-Joshi97,MTJ})
\be
\label{intervals}
\Theta_{u_0}\in {\bf I} =
\left]0,m_3^\frac{3}{2}\left(\frac{13}{3}-\frac{5\sqrt{3}}{2}\right)\right[\cup
\left]m_3^\frac{3}{2}\left(\frac{13}{3}+\frac{5\sqrt{3}}{2}\right),+\infty\right[.
\ee
Now a measure for the interval $\bf R^+\setminus I$ is 
\[
5\sqrt{3}m_3^{\frac{3}{2}}.
\]
It is then clear that the smaller $m_3$ the smaller will be the measure of $\bf R^+\setminus I$
and
consequently, there might exist examples where it is 
easier to find functions $E$ so that $\Theta_{u_0}\in \bf I$. In fact this is the case
for a 
number of known examples (see e.g. \cite{MTJ} for a
graphic illustration). However, this phenomena is not general. The problem 
arises from the fact that 
$\Theta_{u_0}$ depends not only on $m_3$ and $E_2$ but also on the higher order 
coeficients $m_i$ and $E_j$ which although do not appear in the asymptotic ($r\to 0$)
expression 
for $\sigma(t,r)$, can nevertheless be crucial to determine whether $\Theta_{u_0}\in
\bf I$.
This then tells us that the information in the vicinity of the singularity
given by
$E_2,m_3$, which determines the shear rate in (\ref{cinfty}), 
is not enough to determine the nature of the
singularity. It is
therefore crucial to be sufficiently far from the singular region in order to pick up 
information about higher order "inhomogeneous" coeficients which characterise the initial data.
 
We note that Joshi et. al. \cite{JDM} have calculated $\sigma(t,r)$ and $\sigma(0,r)$
for the case $E=0$ and concluded that in approach to the singularity these two
quantities
change with the same rate in $r$. For the case $E\ne 0$, however, one cannot
obtain
expressions for $\sigma(t,r)$ which explicitly depend on $r$ and $t$ so
one cannot compare analytically the rates of change of $\sigma(t,r)$ and
$\sigma(0,r)$ in
the approach to the singularity (as was done for $E=0$). Nevertheless, we prove 
in the next section that the conclusions for $E=0$ do not necessarily carry over to the 
general ($E \neq 0$) case.

We also note that although the function $R$ has
the same asymptotic 
behaviour in the
approach to the singularity for all $E$,
the shear function $\sigma(t,r)$ does not, since the dependence of 
$\sigma(t,r)$ on the function
$E$ persists all the way down to the singularity for the cases
where $E\ne 0$. 
Furthermore, although $R$ "forgets" the $E$ dependence
in the approach to the singularity, the geodesic equations do not and
this is crucial in the study of the nature of the singularities.

To summarize, in this section we have proved that (i) $\sigma(t,r)$ is invariant under the linear 
transformations $\Phi_a$, (ii) $\sigma(t,r)$ diverges in the approach to
the singularity at a rate
depending on $|E|^{3/2}/m$ and (iii) $\sigma(t,r)$ does not 
uniquely specify the nature of the singularity. We have also 
compared our results for $E\ne 0$ to the ones
obtained for $E=0$ by Joshi et
al. \cite{JDM}.
\section{Dependence of the nature of singularity on initial shear $\sigma(0,r)$}
In this section, we study the initial shear $\sigma(0,r)$ as well as its possible
influence on the nature of the
singularities. 
For $t=0$, given $R(0,r)=r$, one can derive
\be 
\label{sigmanot}
\sigma(0,r)=\frac{\sqrt{3}}{3}\left(E+\frac{m}{r}\right)^{-\frac{1}{2}}
\left[\frac{E'}{2}-\frac{E}{r}-\frac{3m}{2r^2}+\frac{m'}{2r}\right].
\ee
The initial radial velocity of collapse is $v_i=\dot R(0,r)$. So, one
can write
\be
\sigma(0,r)=\frac{\sqrt{3}}{3}\left(\frac{v_i}{r}-v_i'\right),
\ee
which implies that $\sigma(0,r)$ increases with $v_i$. This confirms, for the
case of dust, the results of
\cite{Herrera} according to which local anisotropy increases the radial velocity
of collapse.
Now, for $\dot R(t,r)\ne 0$, we have that $\sigma(0,r)=0$ if and
only if \be {\ds
\frac{1}{2}E'-\frac{E}{r}-\frac{3}{2}\frac{m}{r^2}+\frac{1}{2}\frac{m'}{r}}=0.
\ee
\begin{Result}
\label{initials}
Consider a dust spherically symmetric space--time $(M,g)$ with $\sigma(0,r)=0$.
Then the following are equivalent:
\begin{enumerate}
\item $\dot \sigma(0,r)=0$.
\item $(M,g)$ is spatially homogeneous and isotropic.
\item $\sigma (t,r)=0$, for all $t>0$.
\end{enumerate}
\end{Result}
{\bf Proof:} Using (\ref{sigmadot}) for $\sigma(0,r)=0$ we obtain
\be
\dot{\sigma}(t,r)=\frac{\sqrt{3}}{3}\left(\frac{\ddot{R}}{R}-\frac{\ddot{R}'}{R'}\right)
\ee 
which gives 
\be
\dot{\sigma}(0,r)=\frac{\sqrt{3}}{6}\left[\frac{m'}{r^2}-\frac{3m}{r^3}\right]
\ee 
Note that this does not depend on $E$. The condition
$\dot{\sigma}(0,r)=0$ implies $m(r)=m_3r^3$, which in turn from
$\sigma(0,r)=0$ implies $E(r)=E_2r^2$. These correspond to
spatially homogeneous initial data (see e.g. \cite{MTJ}). On the other hand,
spatial homogeneity implies $\dot\sigma(t,r)=0$~$\bullet$ 
\\\\
Result \ref{initials} indicates that, as expected in general, even if we start with
zero initial shear in an
inhomogeneous space--time,
shear will be generated after evolution starts. This can also be seen from the relationship 
between $\dot \sigma(0,r)$ and $\rho(0,r)$:
\be
\label{sigma-rho}
\dot{\sigma}(0,r)=\frac{\sqrt{3}}{6}\left[\rho(0,r)-\frac{3m}{r^3}\right].
\ee
This, however, is not the case
for $E=0$:
\begin{Result}
Consider a dust spherically symmetric space--time with $E=0$. If 
 $\sigma(0,r)=0$ then $\sigma(t,r)=0$.
\end{Result}
{\bf Proof:} For $E=0$, if  $\sigma(0,r)=0$ then one must have
$m=m_3r^3$ which gives $\dot\sigma(0,r)=0$. Now apply Result
(\ref{initials}) ~$\bullet$ 
\\\\
For the case of $E=0$, since shear depends only on $m$ there is a unique correspondence
between
$m$ and $\sigma(0,r)$ and one can therefore substitute the initial free
function $m$ by $\sigma(0,r)$. One
can then study how the initial shear influences the final outcome of collapse, as in
\cite{JDM}.
In particular, if $\sigma(0,r)=0$ the final outcome of collapse is necessarily a
black hole. An interesting question is whether this result necessarily holds for $E\ne 0$.

The situation
is in fact more complicated in the case $E\ne 0$, where as we shall show the same
initial shear
can give rise to different outcomes.
Before doing so we note that one can fine tune the
initial data so that $\sigma(0,r)=0$ and $\dot\sigma(0,r)\ne 0$. 
\begin{Result}
\label{Result-ID}
Consider a dust spherically symmetric collapsing space--time with $m(r)=r^3g(r)$ and
$E(r)=r^2Q(r)$, such that $g$ and $Q$ are $C^2$ real functions in $[0,r_d]$
with $g(0)\ne 0$ and
$Q(0)\ne 0$.
Then
$\sigma(0,r)=0$ if and
only if $Q'+g'=0$.
\end{Result}
{\bf Proof:} Follows directly from expression (\ref{sigmanot})~$\bullet$ 
\\\\
Therefore there can exist situations where 
the initial shear is zero and nevertheless
a naked singularity will be formed.
Consider, for example, an initial data set ${\cal I}$ given by the
functions $m(r)=m_3r^3+m_6r^6$ and $E(r)=E_2r^2+E_5r^5$.
Taking the initial coefficients to be $E_2=m_3=0.1$ and
$E_5=-m_6=1$ we ensure that $\sigma(0,r)=0$ and
$\dot\sigma(0,r)\ne 0$ (this initial data also satisfies the
regularity conditions). Solving  (\ref{quadric-m3m6}) for these
values one finds a positive root $X_0$ which therefore corresponds
to a naked singularity.

Taking the same initial data except for the value of $m_3=0.2$ we find a black
hole, while 
we still have $\sigma(0,r)=0$
and $\dot\sigma(0,r)\ne 0$. One can also find examples for $E_2<0$ such
as $E_2=-0.05,m_3=0.1,E_5=-m_6=1$ giving a
naked singularity and $E_2=-0.05,m_3=1,E_5=-m_6=1$ giving a black hole, both initial sets with
$\sigma(0,r)=0$ and
$\dot{\sigma}(0,r)\ne 0$.

Furthermore one can find open intervals in the coefficients of $m$ and $E$ such
that
the initial shear is zero and yet naked singularities form as in the
next example:
Taking $m(r)=m_3 r^3+m_6 r^6$ and $E(r)=E_2 r^2+E_5 r^5$ with $m_6+E_5=0$, then
from (\ref{theta}) we obtain
\be
\Theta_{u_0}= \frac{E_5}{E_2}\left(\sqrt{1-p_0}+(p_0-\frac{3}{2})G(p_0)\right),
\ee
with $p_0=-E_2/m_3$.
Taking the elliptic case for which $0<p_0\le 1$, i.e.
\be
E_2\in [-m_3, 0 [,
\ee
and consequently $2/3< G(p_0)\le \pi/2$.
In this case $\sqrt{1-p_0}+(p_0-3/2) G(p_0)$ is always bounded and negative. 
So, in order to have $\Theta_{u_0}>0$ and therefore 
avoid shell-crossing we take
$E_5>0$. 
Furthermore
in order to ensure that
$\Theta_{u_0}$ falls in one of the intervals given by (\ref{intervals}), e.g.
$\Theta_{u_0}>m_3^{2/3}(\frac{13}{3}+\frac{5\sqrt{3}}{2})$, we impose
\be
E_5\in \left] \frac{E_2 m_3^{2/3}(\frac{13}{3}+
\frac{5\sqrt{3}}{2})}{\sqrt{1-p_0}+(p_0-3/2)G(p_0)},+\infty\right[.
\ee
So, $\sigma(0,r)=0$ and naked singularities as final state of collapse can be achieved for
open subsets (in the sense made precise above) of the initial data.
The possibility of having the same initial shear with different end states for
the collapse can also
be found
for non--zero initial shear. This can be shown by generalising Result
\ref{Result-ID}:
\begin{Result}
Consider a dust spherically symmetric collapsing space--time with initial data as
given in Result
\ref{Result-ID}. Then two different sets of initial data $\{m^{(1)},E^{(1)}\}$ and
$\{m^{(2)},E^{(2)}\}$
have the same initial shear function $\sigma(0,r)$ if and only if 
$h^{(1)'}+Q^{(1)'}=h^{(2)'}+Q^{(2)'}$.
\end{Result}
An example of this is the naked singularity solution obtained for $i=6$ and $j=5$
with
$m_3=0.1,E_2=-0.05,m_6=-0.9,E_5=1$, and the black hole solution
$m_3=1,E_2=-0.95,m_6=-0.9,E_5=1$ both having the same 
non-zero initial shear profile.
Again one can find open subsets of the initial data which have the same
non--zero initial shear
and yet different outcomes.

To summarize, in this section we have proved that (i) the same initial shear can
result in two different outcomes black holes and naked singularities depending on
$\{m,E\}$,
so initial shear does not uniquely
determine the final state of collapse; (ii) it is possible to have zero initial 
shear and
a naked singularity endstate, and (iii) these results can be obtained for
open sets of initial data.
\section{Dependence of the nature of singularity on inhomogeneity}
The studies of the previous sections for the shear can now be repeated for
inhomogeneity
measures. In this section, we study the influence of the density contrast
on the nature of the singular endstates.
We shall use the pointwise density contrast measure  
\be
\label{dens} 
\rho'(t,r)=
\frac{m''}{R'R^2}-\frac{m'R''}{R^2(R')^2}-\frac{2m'}{R^3}, 
\ee
and start by recalling the following well--known result:
\begin{Result}
Consider a dust spherically symmetric collapsing space--time satisfying the
regularity conditions. If
$\rho'(t,r)=0$ then the final state of collapse is a black hole.
\end{Result}
We shall therefore consider the cases where $\rho'(t,r)\ne 0$.
If we now use the formulae in the Appendix and substitute for $R,R'$ and $R''$ in equation
(\ref{dens}) 
we obtain, after expanding around $\eta=0$,  
\be
\label{rhop}
\rho'(t,r)= -\left(\frac{4 |E|}{m}\right)^3\frac{m'}{\eta^6}+O(\eta^{-3}). 
\ee
Thus as in the case of shear, the inhomogeneity measure $\rho'(t,r)$ depends 
asymptotically on both $E$ and $m$.
In order to prove that  $m'(4 |E|/m)^3$ does not uniquely determine the nature of
the singularity we note that this quantity is invariant under
the linear transformation $\Phi_a(m,E)=(a^{3/2}m,aE)$, for an arbitary free parameter $a>0$. 
In fact, by substituting $R,R'$ and $R''$ in (\ref{dens}) we also find that the all time function 
$\rho'(t,r)$ is itself invariant under $\Phi_a$. 
Interestingly $\Phi_a$ also leaves invariant the
shear function $\sigma(t,r)$ and the examples of section \ref{res} can also be used here
in order to show that the same function $\rho'(t,r)$ can give rise to both naked singularities and
black holes. 


We now study the initial density contrast function $\rho'(0,r)$. From (\ref{dens}) we 
obtain for $t=0$ and $R(0,r)=r$ 
\be
\label{rho-prime}
\rho'(0,r)=\frac{m''}{r^2}-\frac{2m'}{r^3}, 
\ee 
so
$\rho'(0,r)$ does not depend on $E$, which makes it clear that a given 
$\rho'(0,r)$ can correspond to both black holes and
naked singularities, depending on
the choice of $E$
(see e.g \cite{MTJ}). So $\rho'(0,r)$ cannot uniquely determine the
final outcome of collapse, except in the $E=0$ case. 
An interesting question is whether there are naked singularity solutions
for $\rho'(0,r)=0$. In 
order to answer this question we start by recalling a simple result:
\begin{Result}
Consider a dust spherically symmetric space--time with $\rho'(0,r)=0$. Then the
space-time
is spatially homogeneous and isotropic if and only if $\sigma(0,r)=0$.
\end{Result}
{\bf Proof:}
If $\rho'(0,r)=0$ then $m=m_3r^3,~m_3\in{\bf R}\setminus {0}$, in which case
$\sigma(0,r)=0$ if 
and only if $E=E_2r^2$~$\bullet$
\\\\
We recall that in the $E=0$ case the space--time is spatially homogeneous
and isotropic if and only if $\rho'(0,r)=0$. We shall then take the inhomogeneous
$E\ne 0$ cases 
where it might be possible, for $\sigma(0,r)\ne 0$, to find initial data  
such that $\rho'(0,r)=0$ and nevertheless have a naked solution. 
As an example of this we take
$m(r)=m_3r^3,~m_3\in {\bf R^+}$, and $E(r)=E_2r^2+E_5r^5$ such that 
\be
\Theta_{u_0}=\frac{E_5}{E_2}\left(\frac{1}{\sqrt{1-p_0}}-\frac{3}{2}G(p_0)\right ).
\ee 
By taking the elliptic case with $E_2\in {\bf R^-}$ and 
\be
E_5\in \left ] \frac{E_2 m_3^{\frac{3}{2}}(\frac{13}{3}+\frac{5\sqrt{3}}{2})}
{\frac{1}{\sqrt{1-p_0}}-\frac{3}{2}G(p_0)},0 \right [
\ee
we ensure both that $\Theta_{u_0}>0$ and that equation (\ref{general}) has positive roots
corresponding to naked singularities.   

To summarize, in this section we have shown that
(i) $\rho'(t,r)$ is invariant under the linear transformations $\Phi_a$, 
(ii) $\rho'(t,r)$ diverges in the approach to
the singularity at a rate
depending on $m'(|E|/m)^3$, (iii) $\rho'(t,r)$ does not 
uniquely specify the nature of the singularity,
(iv) the same initial
inhomogeneity function
$\rho'(0,r)$ can
result in two different outcomes black holes and naked singularities depending on
$\{m,E\}$,
so $\rho'(0,r)$ does not uniquely
determine the final state of collapse and (v) it is possible to have zero initial 
density contrast and
a naked singularity endstate. These results can be obtained for 
open sets of initial data.
\section{Non-uniqueness, shell-crossing and the initial data set}
\label{initial}
Finally, we consider two important issues in this section: {\bf (I)} Non-uniqueness of space--time extensions 
after shell-crossing singularities and {\bf (II)} Initial data sets based on initial
shear and initial density contrast functions.
\\\\
{\bf (I)} The theme of our discussion has been how certain quantities which
have been suspected of determining the outcome (naked singularity or black hole) of
spherical dust collapse in fact fail to do so. We have emphasised
the significance of the well-known fact that one needs to know
both $E$ and $m$ in order to make this prediction. However there
is another situation in which even the knowledge of both these
functions is not enough to completely determine the space-time;
that is when the initial data are such that shell-crossing
singularities can occur. It has been shown recently how one can
construct a dynamical extension - i.e.\ one based on the field
equations - through a shell-crossing singularity and obtain global
weak solutions of the field equations for $R>0, t>0$
\cite{nolan03}. However these weak solutions are not unique to the
future of the shell-crossing. On the other hand, the earliest
point of the shell-crossing is always globally naked, regardless of
how the extension is constructed, so the non-uniqueness is not so
severe as to allow the same initial data give rise to either a
black hole or a naked singularity.
\\\\
{\bf (II)} As we have seen, neither $\sigma(0,r)$ nor $\rho'(0,r)$ alone can
uniquely determine the final outcome of collapse. However, both functions
might form a perfectly good initial data set.
The question is how from a set of observables
${\cal J}=\{\rho'(0,r),\sigma(0,r)\}$, 
given on the onset of collapse, can one predict (uniquely) the final outcome of
collapse.

Now $m(r)$ is
uniquely determined by $\rho'(0,r)$ through the linear ODE (\ref{rho-prime}):
\be
\label{unique-m}
m''-\frac{2}{r}m'=r^2\rho'(0,r),
\ee
with initial conditions $m(0)=m'(0)=0$.
The shear $\sigma(0,r)$ can be written as a
function of a new variable $w(r)=E+m/r$: 
\be
\sigma(0,r)=\frac{\sqrt{3}}{3}w^{-\frac{1}{2}}(-\frac{w}{r}+\frac{1}{2}w'),
\ee
with initial conditions $w(0)=w'(0)=0$.
So, from $\sigma(0,r)$ one can determine $w(r)$ by solving the linear ODE
\be
\label{ode1}
y'-\frac{2}{r}y=\sqrt{3}\sigma(0,r),
\ee
where $w=y^2$. So, using (\ref{unique-m}), we can get $m(r)$ from $\rho'(0,r)$.
Knowing $m(r)$ and $\sigma(0,r)$ we can determine $E$ from
(\ref{ode1}).
So, if we know both the
density contrast $\rho^\prime(0,r)$ and the anisotropy measure
$\sigma(0,r)$ we can recover both $E$ and $m$, and so predict the
outcome of collapse. Since this process relies on solving linear ODEs, then one
can ensure
that solutions exist and are
unique, for the given set of initial conditions.

It is worthwhile noting that even though the set $\{\sigma (0,r), \rho ' (0,r)\}$
can specify the initial data uniquely, the set $\{\sigma (t,r), \rho' (t,r)\}$ specified
at any other time cannot. The reason is that the latter is
invariant under the one-parameter set of linear transformations $\Phi_a$
acting on the initial data set $\{m,E\}$, while the former is not and 
this is because even though initially one can
use the rescaling to define $R(0,r) = r$, 
this cannot be repeated at subsequent times. 
\section{Conclusions}
We have considered LTB collapsing models in order to study the influence of inhomogeneity
and anisotropy 
in the nature of the resulting singularities.
We have 
studied the shear and density contrast
in three ways: (i) by considering the all time functions $\sigma(t,r)$ and $\rho'(t,r)$, 
(ii) by considering the asymptotic functions $\sigma(t,r)$ and $\rho'(t,r)$ and (iii) by considering the
initial functions $\sigma(0,r)$ and $\rho'(0,r)$. 

We have found that both shear $\sigma(t,r)$ and density contrast $\rho'(t,r)$ 
are invariant under 
a one parameter set of linear transformations
acting on the initial data set $\{m,E\}$. We have also found that asymptotically 
one cannot establish, in general, a direct link between the strenght of shear 
(or density constrast) and the nature of the singularities.
Finally we have found that
the same initial shear (or initial density contrast) 
can give rise to both naked singularities and black holes, depending on the choice
of the initial density contrast (or initial shear, respectively).
In particular
this can happen for zero initial shear
or zero initial density contrast. 
These results were obtained for open sets of initial data. 
We conclude that neither anisotropy nor inhomogeneity features 
as given by the shear or density contrast can alone 
uniquely specify the end state of collapse for $E\ne 0$, 
in contrast to the results established for $E=0$ in \cite{JDM}.

However, we have proved that if we 
know both the initial shear $\sigma(0,r)$ and the initial density 
contrast $\rho'(0,r)$
functions
then, given the appropriate initial conditions, we
can
determine uniquely the initial LTB data functions $\{m,E\}$ and 
hence determine the nature of singularity.
In this
sense
one needs information about both the inhomogeneity and anisotropy at the initial
hypersurface 
in order to predict the outcome of the collapse. 
This is important for understanding the effects of the
initial physical state
in determiming the end state
of gravitational collapse.



\vspace{.2in}

\centerline{\bf Acknowledgments}
\vspace{.2in}
FCM thanks S\'ergio Gon\c{c}alves for interesting discussions, British
Council/CRUP for grant N$13/03$, Funda\c{c}\~ao Calouste Gulbenkian for
grant 21-58348-B and Centro de Matem\'atica, Universidade do Minho, for support.

\section{APPENDIX}
\subsection{$E<0$}
\be 
\frac{\partial \eta}{\partial
t}=\frac{2(-E)^{3/2}}{m}\frac{1}{\cos\eta-1}<0. 
\ee
\be 
\frac{\partial\eta}{\partial
r}=\frac{1}{1-\cos\eta}\left[\frac{m}{2(-E)^{\frac{3}{2}}}
\left(\frac{2(-E)^{\frac{3}{2}}}{m}\right)'(\eta-\sin\eta)+\frac{2(-E)^{\frac{2}{3}}}{m}t_c'\right].
\ee

\be
R''=\left(\frac{m}{2(-E)}\right)''(1-\cos\eta)+2\left(\frac{m}{2(-E)}\right)'\sin\eta\frac{\partial\eta}{\partial
r}+\frac{m}{2(-E)}\left(\cos\eta\frac{\partial\eta}{\partial
r}+\sin\eta\frac{\partial^2\eta}{\partial r^2}\right).
\ee

\subsection{$E>0$}
\be
\frac{\partial \eta}{\partial
t}=-\frac{2E^{\frac{3}{2}}}{m}\frac{1}{\cosh\eta-1}<0.
\ee
\be
\frac{\partial \eta}{\partial r}=\frac{1}{\cosh\eta-1}
\left[\frac{m}{2E^{3/2}}\left(\frac{2E^{\frac{3}{2}}}{m}\right)'(\sinh\eta-\eta)+\frac{2E^{\frac{3}{2}}}{m}t_c'
\right].
\ee
\be
R''=\left(\frac{m}{2E}\right)''(\cosh\eta-1)+2\left(\frac{m}{2E}\right)'\sinh\eta\frac{\partial\eta}{\partial
r}+\frac{m}{2E}\left(\cosh\eta\frac{\partial\eta}{\partial
r}+\sinh\eta\frac{\partial^2\eta}{\partial r^2}\right).
\ee

\end{document}